\documentclass[aps,prd,groupedaddress,nofootinbib,letterpaper]{revtex4}

\usepackage{amsmath}
\usepackage{amsfonts}
\usepackage{amssymb}
\usepackage{amsthm}

\usepackage{braket,mathtools}
\usepackage{mathrsfs}
\usepackage{stmaryrd}

\usepackage{hyperref}
\usepackage{color}

\newcommand{\cala}{{\cal A}}
\newcommand{\call}{{\cal L}}
\newcommand{\calo}{{\cal O}}
\newcommand{\calh}{{\cal H}}

\newcommand{\beq}{\begin{equation}}
\newcommand{\eeq}{\end{equation}}
\newcommand{\bea}{\begin{eqnarray}}
\newcommand{\eea}{\end{eqnarray}}
\newcommand{\hf}{\frac{1}{2}}

\newcommand{\xb}{{\bar x}}

\begin{document}
\begin{titlepage}

\title{Gravitational splitting at first order:\\  Quantum information localization in gravity}

\author{William Donnelly}
\email{wdonnelly@perimeterinstitute.ca}
\affiliation{Department of Physics, University of California, Santa Barbara, CA 93106}
\affiliation{Perimeter Institute for Theoretical Physics,
31 Caroline Street North, Waterloo, Ontario N2L 2Y5, Canada\footnote{Current affiliation}}

\author{Steven B. Giddings}
\email{giddings@ucsb.edu}
\affiliation{Department of Physics, University of California, Santa Barbara, CA 93106}

\begin{abstract}
We explore the important fundamental question of how quantum information is localized in quantum gravity, in a perturbative approach.  Familiar descriptions of localization of information, such as via tensor factorization of the Hilbert space or a net of commuting subalgebras of operators, conflict with  basic gravitational properties -- specifically gauge invariance -- already at  leading order in perturbation theory.  However, previous work found that information can be classically localized in a region in a way such that measurements, including those of the gravitational field, outside the region are insensitive to that information, and only measure total Poincar\'e charges.  This paper shows that, working to leading order in the gravitational coupling, a similar quantum result holds, leading to a definition of a ``gravitational splitting"  on the Hilbert space for gravity.  Such localization of information also argues against a role for ``soft hair" in resolving the information problem for black holes.  This basic mathematical structure plausibly plays a foundational role in the quantum description of gravity.

\end{abstract}

\maketitle

\end{titlepage}

\section{Introduction}

A longstanding question playing an important role in various approaches to quantum gravity is that of how quantum information is localized.   For example, the idea that information can be equivalently represented in a dual ``boundary" theory is central to the conjectured AdS/CFT equivalence, and the idea that information is accessible through the gravitational field has been suggested\cite{Hawk,HPS1,HPS2} as a possible resolution to the information problem for black holes.  This question of localization of information in a gravitational theory arises because the usual notion of localization from local quantum field theory (LQFT) does not respect the gauge symmetries of gravity, which at least in the geometrical approximation correspond to diffeomorphisms.  Put simply: how do we construct ``qubits" in gravitational theories, and are they in any sense localized?  

Specifically, local operators of LQFT are not gauge (diffeomorphism) invariant.    There are various approaches to constructing operators that are gauge invariant.  One involves constructing integrals over all of spacetime of products of local operators, and is in a generic state maximally nonlocal; this type of construction is reviewed for example in \cite{GMH}.  Another approach is to start with a local operator of LQFT, and ``gravitationally dress" it to make it gauge invariant.  Explicit such constructions appear in \cite{DoGi1}, and related earlier constructions are those of \cite{Heem,KaLigrav}.  These operators create not only particles, but also their gravitational fields, reflecting the statement that a particle is inseparable from its gravitational field.  In particular, these operators {\it must} have nontrivial support extending to infinity, as shown in \cite{DoGi2}, so are also nonlocal, although
if one works perturbatively in the gravitational coupling $\kappa=\sqrt{32 \pi G}$ the nonlocality is small in $\kappa$.  But this raises the important question regarding whether in gravitational theory there is any precise notion of localized quantum information,\footnote{For related discussion see \cite{Jaco}.} or whether any information is necessarily delocalized in the collective state of a matter distribution and its gravitational field.

In fact, this question appears likely to play a fundamental role in construction of a theory of quantum gravity.
  This is because the notion of a ``subsystem," localizing information, is usually part of the fundamental structure assumed in a quantum theory.  In finite systems, or locally finite ones such as a lattice, this localization arises from a tensor factorization of the Hilbert space.  In LQFT, such localization is also a fundamental starting principle\cite{Haag}, encoded in the statement that subalgebras of operators localized to spacelike separated regions commute.  It is both interesting and apparently important that these statements appear to be significantly modified in gravity.  Plausibly a good approach towards understanding quantum gravity is
to take a ``quantum-first" viewpoint that it should respect basic axioms of quantum mechanics, particularly the existence of a Hilbert space;  then, a large 
part of the problem of formulating the theory is  to find appropriate mathematical structure on that Hilbert space for describing gravity\cite{QFG,QGQF}.
This is guided partly by a correspondence principle for weak gravitational fields.  
In such an approach, it seems that a basic element of the mathematical structure is plausibly such a notion of subsystems, just as this notion enters at the foundational level for LQFT and other quantum systems.

A key point appears to be that while one is required to nonlocally gravitationally dress local operators, there is significant latitude in {\it how} an operator is dressed.  This led to description  in \cite{DoGi3} of how one could begin with a classical matter configuration in a given region and 
find a corresponding {\it classical} gravitational field such that gravitational (or other) observations outside that region could not detect the details of the configuration.  This (and a related construction in QED) suggested the idea of a ``gravitational splitting,"\footnote{This was called split structure in \cite{DoGi3}.} which is roughly a Hilbert subspace of states that is indistinguishable by measurements outside a region.  This paper will focus on the question of construction of such gravitational splittings in  the {\it quantum} theory of matter plus gravity, working to leading order in $\kappa$.  Construction of such splittings demonstrates a way to localize quantum information at this order.  It also appears to lend weight to arguments\cite{DoGi3} that soft hair does not offer a way to determine the information content of a region, which plausibly extend to the case of black hole interiors.

\section{The question of localizing quantum information}

The present goal is to understand the extent to which information can be ``localized" in a gravitational theory, at least perturbatively.  We first recall how information is localized in non-gravitational local quantum field theory (LQFT).

In finite or locally-finite (such as a lattice) quantum systems, information is localized by providing a tensor factorization of the Hilbert space, {\it e.g.} 
\beq\label{prodspace}
\calh=\bigotimes_i \calh_i\ ,
\eeq
where $\calh_i$ are a collection of Hilbert spaces.  Then, states can be independently excited in the different tensor factors of the Hilbert space.  However, the Hilbert space of LQFT does not have such a structure, due to the generic von Neumann type-III behavior of its operator algebras.\footnote{See \cite{BuVe}, and other references in \cite{YngR}.}  This means that if we for example divide space into two at a surface, the vacuum $|0\rangle$ of LQFT has infinite entanglement between excitations on the two sides of the division, and simple product states are in a different Hilbert space than $|0\rangle$.  

However, in LQFT we can describe localized information through the existence of subalgebras of operators associated with different regions.  If $U_1$ and $U_2$ are two spacelike-separated regions, then operators compactly supported in these different regions form commuting subalgebras.  These operators can be thought of as creating independent excitations in the two regions.  In general, the collection, or net, of such subalgebras mirrors the open set structure (topology) of the spacetime manifold\cite{Haag}, and effectively defines a subsystem structure.

Indeed, one can go a step closer to \eqref{prodspace}, and make it clear that independent information can exist ``in a region" in LQFT.  This arises from the notion of a split vacuum.  Let $U$ be a neighborhood, and $U_\epsilon$ be an $\epsilon$-extended open set containing $U$.  Denote the subalgebra associated to $U$ as $\cala_U$, and that associated to the complement $U_\epsilon'$ as $\cala_{U_\epsilon'}$.  Then, it has been shown (see \cite{Haag}, and references therein) that there exists a split vacuum 
$|U_\epsilon\rangle$ so that for any $A\in \cala_U$ and $A'\in U_\epsilon'$, 
\beq\label{splitst}
\langle U_\epsilon| A A'|U_\epsilon\rangle= \langle0| A|0\rangle \langle0| A'|0\rangle\ :
\eeq
observations outside $U_\epsilon$ are incapable of distinguishing operation of different operators inside $U$.  So, for example, we could pick a collection of operators $A_I\in\cala_U$, and form the states
\beq\label{qudits}
|\psi_I\rangle = A_I |U_\epsilon\rangle\ ;
\eeq
observations using operators $A'\in \cala_{U_\epsilon'}$ cannot distinguish such states, and so these states can be thought of as describing localized information -- if $I=1,2,$ a localized qubit.

We next turn to the analogous question for a gravitational theory, taken for simplicity to be that of a mass-$m$ scalar $\phi$ minimally-coupled to gravity, with lagrangian
\beq\label{scalarlag}
\call = \frac{2}{\kappa^2} R -\hf\left[(\nabla\phi)^2+m^2\phi^2\right]\ .
\eeq
We work perturbatively about flat space, although these considerations generalize; for example the AdS context is described in \cite{GiKi}.
The essential problem is that $\phi(x)$ is no longer an observable: it is not gauge invariant under diffeomorphisms, which act infinitesimally with parameters 
$\xi^\mu(x)$ as $\delta x^\mu=\kappa\xi^\mu$ and
\beq
\delta_\xi \phi(x) = -\kappa \xi^\mu\partial_\mu\phi(x) .
\eeq
Working perturbatively in $\kappa$, \cite{DoGi1} showed that one may ``dress" $\phi$ to give gauge invariant operators $\Phi(x)$, but these must\cite{DoGi2} involve integrals to infinity of the metric perturbation $h_{\mu\nu}(x)$, defined by
\beq\label{metexpan}
g_{\mu\nu}=\eta_{\mu\nu} + \kappa h_{\mu\nu}\ ,
\eeq
 and thus obey a nonlocal algebra\cite{SGalg,DoGi1}.    

In essence, as noted, an operator that creates a particle must also create its gravitational field.  This complicates the question of localization of information in gravity, since one then expects that information about the particle state is contained in the gravitational field, and thus may be measured far from the particle.  But the question is {\it how much} information is contained in the gravitational field, and specifically whether there is localized information that is not measurable via the gravitational field outside a region.

If gravity permits global symmetries, there is in fact a trivial example of localization of information.  Suppose that $\phi_1(x)$ and $\phi_2(x)$ are two fields related by such a global symmetry, and consider corresponding identically-dressed operators $\Phi_1(x)$ and $\Phi_2(x)$.  Then, there is no way to asymptotically perturbatively distinguish\footnote{For some discussion of the nonperturbative situation, see \cite{DoGi3}.} states created by $\Phi_1(x)$ or $\Phi_2(x)$: these operators can be used to create ``localized gravitational qubits," analogous to \eqref{qudits}.  However, the existence of such global symmetries in a gravitational theory has long been questioned, either because of black hole evaporation, spacetime wormholes\cite{GiSt}, or the nature of symmetries in string theory\cite{BaDi}.   While the ultimate role of global symmetries in quantum gravity is unclear, 
we seek to examine the more general question of localization of information, without such a symmetry.

\section{Gravitational dressing of operators and states}

We first need to revisit and extend the discussion of gravitational dressing given in \cite{DoGi1}.  There, it was explicitly found that to leading order in $\kappa$, the LQFT operator $\phi(x)$ could be promoted to a diffeomorphism-invariant operator $\Phi(x)$.  This was done by finding a dressing $V^\mu(x)$  linear in $h_{\mu\nu}$ and transforming as 
\beq\label{Vdiff}
\delta_\xi V^\mu(x) = \kappa \xi^\mu(x)\ 
\eeq
under an infinitesimal diffeomorphism, which to leading order in $\kappa$ acts as
\beq\label{hdiff}
\delta_\xi h_{\mu\nu}=-\partial_\mu\xi_\nu -\partial_\nu\xi_\mu\ .
\eeq
Then, 
\beq
\Phi(x)=\phi(x^\mu+V^\mu(x)) 
\eeq
is diffeomorphism invariant to $\calo(\kappa)$.  The dressing $V^\mu(x)$ is not unique; different choices exist, including the gravitational line and Coulomb dressings investigated in \cite{DoGi1}.

To promote a more general LQFT operator $A$ to a diffeomorphism-invariant version $\hat A$, note that the diffeomorphisms are generated by the constraints
\beq\label{constraints}
{\cal C}_\mu = \left(\kappa T_{\mu\nu} - \frac{4}{\kappa} G_{\mu\nu}\right)n^\nu\ ,
\eeq
with $T_{\mu\nu}$ the stress tensor, $G_{\mu\nu}$ the Einstein tensor, and $n^\mu=(1,0,0,0)$ the unit timelike vector.   This may for example be seen from the covariant canonical approach\cite{AsMa,ABK,CrWi,Crnk,Zuck,LeWa,IyWa}, as reviewed in \cite{DoGi2,GiKi}.  
So, a diffeomorphism-invariant observable $\hat A$ should commute with ${\cal C}_\mu$,
\beq\label{constcomm}
[{\cal C}_\mu (x), {\hat A} ]=0\ .
\eeq

The approach taken will be to solve the condition \eqref{constcomm} perturbatively in an expansion in $\kappa$, starting with the $\kappa=0$ operator $A$ of LQFT.
To begin, the expansion of the metric \eqref{metexpan} gives
\beq
G_{\mu\nu} = \kappa G^{(1)}_{\mu\nu} - \frac{\kappa^2}{4} t_{\mu\nu}\ ,
\eeq
where $G^{(1)}_{\mu\nu}$ is linear in $h$, and $t_{\mu\nu}$ is an effective stress tensor that is quadratic and higher order in $h$.  Then the constraints \eqref{constraints} become
\beq\label{constraintsT}
{\cal C}_\mu = c_\mu +\kappa \left( T_{\mu\nu} +t_{\mu\nu} \right)n^\nu\ ,
\eeq
where $c_\mu=-4 G^{(1)}_{\mu\nu}n^\nu$.  Explicitly, 
\beq
c_0= -2(\partial_i \partial_j h_{ij} - \partial_i \partial_i h_{jj})\quad,\quad c_i = -2(\partial_j \dot{h}_{ij} - \partial_i \dot{h}_{jj} + \partial_i \partial_j h_{0j} - \partial_j \partial_j h_{0i}) \ .
\eeq
Equal-time commutators of  $c_\mu$ with the  $h_{\mu\nu}$ generate the linearized diffeomorphisms \eqref{hdiff}, and correspondingly, as seen in \eqref{Vdiff}, a dressing $V^\mu(x)$ should satisfy
\beq\label{Vcomm}
[c_\mu(t,\vec x),V^\nu(t,\vec x')]= i\kappa\delta_\mu^\nu \delta^3(\vec x-\vec x')\ .
\eeq
This may be explicitly checked, {\it e.g.} using a covariant gauge fixing as in appendix B of \cite{DoGi1}.  
Then, given a dressing $V_\mu$ satisfying \eqref{Vdiff}, \eqref{Vcomm}, the invariance condition \eqref{constcomm} is easily seen to be solved to $\calo(\kappa)$ by
\beq\label{dressop}
\hat A= A+i\int d^3x V^\mu(x) [T_{0\mu}(x),A] +\calo(\kappa^2)= e^{i\int d^3x V^\mu(x) T_{0\mu}(x)}Ae^{-i\int d^3x V^\mu(x) T_{0\mu}(x)}+\calo(\kappa^2)\ 
\eeq
(generalizing eq. (33) of \cite{DoGi1}).

To understand whether states can carry local information, one also needs the dressed version of a state such as $|\psi_I\rangle=A_I|U_\epsilon\rangle$.  Eq. \eqref{dressop} suggests
\beq\label{dressst}
|\widehat\psi_I\rangle = \left[1+ i\int d^3x V^\mu(x) T_{0\mu}(x)\right]|\psi_I\rangle +\calo(\kappa^2)
\eeq
as the analogous dressed state.  Here it suffices to assume that the state is built  on a vacuum $|U_\epsilon\rangle_\phi$ which 
is split just for $\phi$; for $h$ this is the usual vacuum.  As in Gupta-Bleuler quantization of gauge theory, we do not require ${\cal C}_\mu(x)$ to annihilate dressed physical states\cite{Gupt}, but only that it have vanishing matrix elements between such states,
\beq\label{pstatecond}
\langle\widehat\psi_I|{\cal C}_\mu(x)|\widehat\psi_J\rangle=0\ .
\eeq
If one assumes that the positive-frequency part of $c_\mu$ annihilates the vacuum at $\calo(\kappa^0)$,
\beq
c_\mu^{(+)}(x)|U_\epsilon\rangle_\phi=0\ ,
\eeq
then \eqref{pstatecond} follows to order $\kappa$ from the definition \eqref{dressst} and from \eqref{Vcomm}.

We could alternately try to introduce a split vacuum for $h_{\mu\nu}$, and seek to  enforce the vanishing of the constraints \eqref{pstatecond} via an expression similar to 
\eqref{dressst}, involving $t_{\mu\nu}$.    However, the gauge non-invariance of $t_{\mu\nu}$ is an added complication.  While this  would be useful if we wanted to consider states corresponding to dressed gravitons in $U$, this construction will not be needed here and will be deferred to future work.

\section{First-order gravitational splitting}

Given the preceding constructions, and particularly the flexibility in choosing the dressing $V^\mu$, the next question is whether one can choose a dressing so that different excitations localized to a neighborhood $U$ can be dressed in a way so they are indistinguishable outside the extended neighborhood $U_\epsilon$.  
Of course, all such excited states $|\psi_I\rangle$ yield a 
non-trivial gravitational field outside $U_\epsilon$, so the question is in what circumstances this dressing is insensitive to the state.

Specifically, since we are working to $\calo(\kappa)$, consider the matrix element
\beq\label{hmat}
\langle\widehat\psi_I |h_{\mu\nu}(\xb) |\widehat\psi_J\rangle
\eeq
between two states of the form \eqref{dressst}.  In a more complete treatment, we really want to use a diffeomorphism-invariant, dressed, version of $h_{\mu\nu}(\xb)$, but to leading order in $\kappa$, \eqref{hmat} suffices.  Ultimately, we are also interested in higher-point functions, {\it e.g.} $\langle\widehat\psi_I |h_{\mu\nu}(\xb) h_{\lambda\sigma}({\bar y})|\widehat\psi_J\rangle$, {\it etc.}, but these involve the $\calo(\kappa^2)$ corrections that have been neglected.  So, we will check insensitivity of \eqref{hmat} to the state only to first order in $\kappa$.  

For $\xb\in U_\epsilon'$, $\langle\psi_I |h_{\mu\nu}(\xb) |\psi_J\rangle=0$, and one finds
\bea\label{hmatint}
\langle\widehat\psi_I |h_{\mu\nu}(\xb) |\widehat\psi_J\rangle&=& -i\langle\psi_I |\int d^3x T_{0\lambda}(x) \left[V^\lambda(x),h_{\mu\nu}(\xb)\right]|\psi_J\rangle + \calo(\kappa^2)\cr
&=& -i\int d^3x \left[V^\lambda(x),h_{\mu\nu}(\xb)\right] \langle\psi_I |T_{0\lambda}(x)|\psi_J\rangle+\calo(\kappa^2)\ ,
\eea
where the second line uses the linearity of $V$ in $h$, so that $[V,h]$ is a c-number.

The next goal is to find a dressing $V^\mu(x)$ that is ``as insensitive as possible" to the details of the matter distribution inside $U$.  This is done by first picking a point $y\in U$.  Then, we construct the dressing by ``dressing the point $y$" with some  ``standard" dressing $V_S^\mu(y)$, and then adding dressing connecting the point $y$ to a more general point $x\in U_\epsilon$.  Specifically, consider the expression\cite{QGQF}
\beq
V_{L\mu}(x,y)=-\frac{\kappa}{2}\int_y^x dx^{\prime\nu}\left\{h_{\mu\nu}(x') - \int_y^{x'}dx^{\prime\prime\lambda}\left[\partial_\mu h_{\nu\lambda}(x'')-\partial_\nu h_{\mu\lambda}(x'')\right]\right\}\ .
\eeq
Given the transformation \eqref{hdiff}, one may easily check that
\beq\label{Vldiff}
\delta_\xi V_{L\mu}(x,y) =\kappa \left\{ \xi_\mu(x)-\xi_\mu(y)+ \hf(x^\nu-y^\nu)\left[\partial_\mu\xi_\nu(y)-\partial_\nu\xi_\mu(y)\right]\right\}\ .
\eeq
Next, pick a particular standard dressing $V_S^\mu(y)$, satisfying \eqref{Vdiff}; this could, for example, be either the gravitational line  or Coulomb dressings of \cite{DoGi1}.  Then, combining \eqref{Vdiff} and \eqref{Vldiff} shows that
\beq\label{Vsplit}
V^\mu(x) =  V_L^\mu(x,y) + V_S^\mu(y) +\hf (x-y)_\nu\left[\partial^\nu V_S^\mu(y)-\partial^\mu V_S^\nu(y)\right]
\eeq
satisfies the correct transformation law\eqref{Vdiff}.

The dressing \eqref{Vsplit} is just what is needed for maximum insensitivity to the details of the state.  To see this, note that by the split property \eqref{splitst}, the integrand of \eqref{hmatint} vanishes for $x$ outside $U_\epsilon$, and this means the term involving $V_L$ doesn't contribute.  As a result, one finds
\beq\label{hmatres}
\langle\widehat\psi_I |h_{\mu\nu}(\xb) |\widehat\psi_J\rangle = {\tilde h}^{S\lambda}_{\mu\nu}(\xb,y) \langle\psi_I |P_\lambda |\psi_J\rangle + \hf\partial_y^{\lambda}{\tilde h}^{S\sigma}_{\mu\nu}(\xb,y) \langle\psi_I | M_{\lambda\sigma}|\psi_J\rangle +\calo(\kappa^2)
\eeq
where we have defined a collection of standard dressing fields, labelled by $\lambda$, 
\beq
{\tilde h}^{S\lambda}_{\mu\nu}(\xb,y) = -i[h_{\mu\nu}(\xb),V^\lambda_S(y)]\ ,
\eeq
and where $P_\mu$ and $M_{\mu\nu}$ are the total momenta and angular momenta operators,
\beq
P_\mu=-\int d^3x T_{0\mu}(x)\quad ,\quad M_{\mu\nu}= - \int d^3x \left[(x-y)_\mu T_{0\nu}(x)-(x-y)_\nu T_{0\mu}(x)\right]\ .
\eeq
In short, the matrix elements of the metric outside $U_\epsilon$ only depend on the matrix elements of the Poincar\'e charges.  Of course no localized state can be an eigenstate of all the Poincar\'e charges, but any subspace of states where these matrix elements take a given fixed value produces identical matrix elements for $h_{\mu\nu}$ outside the neighborhood.  

Thus a subspace of such states localized to $U_\epsilon$ can encode information not accessible by $\calo(\kappa)$ measurements of the gravitational field $h_{\mu\nu}$ outside.  A {\it gravitational splitting} is defined as\cite{DoGi3}\footnote{In \cite{DoGi3} this was called a ``split structure."} a collection of Hilbert subspaces $\calh^i_{U_\epsilon} \subset \calh$ so that for any two states $|\psi\rangle$, $|\psi'\rangle\in \calh^i_{U_\epsilon}$, and any operator $\bar A$ localized outside $U_\epsilon$, 
\beq
\langle \psi|\bar A|\psi'\rangle = \langle \psi|\psi'\rangle \langle i|\bar A|i\rangle\ ,
\eeq
so the value of $\bar A$ is independent of the choice of state within $\calh^i_{U_\epsilon}$.  The preceding construction thus yields a gravitational splitting to leading order $\kappa$, with the labels $i$ given by the values of the Poincar\'e charges in the individual subspaces.  This generalizes the classical result of \cite{DoGi3}, and implies the existence of gravitational dressing such that the quantum information contained in a matter configuration, aside from its Poincar\'e charges, can be localized in a neighborhood, to this order.  An important question is generalizing this construction to higher order.

Given a non-gravitational LQFT state localized to a neighborhood, the preceding construction gives a dressing of it that, outside the enlarged neighborhood, creates initial data for a gravitational field determined by the Poincar\'e charges and the choice of standard field.   This can be thought of as arising from adding a radiation field (solution of the source-free equations) to a given gravitational field configuration to put it in standard form.  As noted, examples are a gravitational line dressing field generalizing that found in \cite{DoGi1}, or the generalization of the Schwarzschild dressing field found in \cite{DoGi1}, giving in general a linearized boosted Kerr field.  The existence of such field configurations connects to results for the full nonlinear theory in the literature.  In particular, \cite{CaSc} (also see \cite{Chru}) showed that initial data for the fully nonlinear gravitational field arising from a localized matter distribution may be restricted to a cone going to infinity, generalizing the gravitational line, and \cite{CoSc} showed that the initial data may be chosen to be boosted Kerr outside a neighborhood, albeit in the special case of a solution of the vacuum equations.  Of course, in general as these fields evolve forward, they will produce outgoing radiation. For example, the gravitational line initial data produces radiation as it settles down to a Coulomb configuration.  

The preceding results also appear pertinent to the question of whether soft hair\cite{HPS1,HPS2} can help with the problem of unitarizing black hole evolution.  Specifically, given a region $U$, these results show that initial data may be constructed for the gravitational field such that outside $U_\epsilon$, the field only depends on the Poincar\'e charges of a matter distribution inside $U$.  The values of the soft charges will depend on which standard dressing is chosen outside $U_\epsilon$, but not on the other details of the state inside $U$.  Moreover, other values of the soft charges may be found by superposing a homogeneous (sourceless) solution of the linearized gravitational field equations on whatever standard dressing has been chosen.\footnote{The basic soft-charge analysis is given in terms of linearized perturbations, so higher-order effects are not considered.}  So, while there is a large amount of extra information in the soft charges, it is uncorrelated with the information in $U$; the former information parameterizes the additional gravitational radiation that has been added on top of the underlying configuration.\footnote{For related arguments, see \cite{MiPo,BoPo}.}  While the preceding analysis has been carried out in a flat background, one certainly expects that if there is such ``localized" information that is invisible to soft charges outside a general region, there is likewise localized information inside a black hole.  Work is in progress to further describe properties of dressing in black hole backgrounds\cite{GiWe}.

The preceding discussion also suggests a diminished role for higher multipoles of the gravitational field in the AdS/CFT correspondence.  It has been argued that since all multipoles fall at the same rate in AdS, that could play an important role in a holographic correspondence\cite{HoHu,FKM}.  However, similar dressing constructions have been given in AdS\cite{GiKi}, and plausibly can be used to argue that higher multipoles can be likewise removed by choice of radiation field.

Finally, as noted in the introduction, gravitational splittings or related concepts may play an important role in the foundational structure of a theory of quantum gravity.  Starting with Hilbert space, one  needs to find a mathematical structure  on it that reproduces spacetime structure in the weak-gravity correspondence limit.  One approach to specification of such a ``gravitational substrate"\cite{QFG} is plausibly in terms of a network of Hilbert spaces, such as arising from gravitational splittings, and related by inclusion.

\section{Acknowledgements}

The work of SG was supported in part by the U.S. DOE under Contract No. 
{DE-SC}0011702, and that of WD was supported in part by the University of California and in part by Perimeter Institute for Theoretical Physics. Research at Perimeter Institute is supported by the Government of Canada through Innovation, Science and Economic Development Canada and by the Province of Ontario through the Ministry of Research, Innovation and Science.

\bibliographystyle{utphys}
\bibliography{gravsplit}

\end{document}